\documentclass[10pt,onecolumn]{article}
\newcommand{\be}[1]{\begin{equation}\label{#1}}
\newcommand{\ee}{\end{equation}}

\newcommand{\Tr}{\mbox{Tr}}

\newcommand{\R}{\mbox{I}\!\mbox{R}}
\newcommand{\et}{{\em et al }}
\usepackage{amsmath}
\usepackage{amsfonts}
\usepackage{amssymb}
\usepackage[dvips]{graphicx}
\usepackage[margin=3cm]{geometry}
\begin{document}

\title{From quantum graphs to quantum random walks}
\author{Gregor Tanner\\
{\small School of Mathematical Sciences, University of Nottingham, UK }}
\maketitle
 
\begin{abstract}
We give a short overview over recent developments on quantum graphs and
outline the connection between general quantum graphs and so-called quantum
random walks.

\end{abstract}

\section{Introduction}
The study of quantum mechanics on graphs has become an important tool
for investigating the influence of classical dynamics on
spectra, wavefunctions and transport properties of quantum systems.
Quantum networks have been used with great success to model quantum phenomena
observed in disordered metals and mesoscopic systems (Chalker and Coddington 
1988); typical behaviour found in diffusive systems such as 
localisation - delocalisation transitions (Freche \et 1999), 
transport properties (Pascaud and Montambaux 1999) and quantum spectral
statistics (Klesse and Metzler 1997) have been studied on graphs in the limit
of infinite network size. Kottos and Smilansky (1997, 1999) looked at
quantum graph models for general, non-diffusive graphs; this approach was
motivated by trying to understand the validity of the Bohigas-Giannoni-Schmit 
(BGS) conjecture (Bohigas \et 1984) in terms of periodic orbit trace formula.
The conjecture relates the properties of the classical dynamics of a systems
to the spectral statistics of its quantum counterpart and states that for 
chaotic systems the statistics depends only on the symmetries of the problem
and follows random matrix theory (RMT) otherwise.

Making use of the fact that periodic orbit trace formula are exact on 
quantum graphs and that there are only a finite number of different length
scales on a finite graph, a series of remarkable results have been obtained 
over the last couple of years.  A closed form quantisation conditions in 
terms of periodic orbits has been given by Bl\"umel \et (2002); Barra and 
Gaspard (2000) derived an integral expression for the level spacing 
distribution starting form the periodic orbit trace formula. Furthermore,
Schanz and Smilansky (2000) described localisation on one-dimensional chains 
in terms of combinatorial expression using periodic orbits. The most far 
reaching development is maybe due to Berkolaiko \et (2002,2003), who, 
inspired by work from Sieber (2002) and Sieber and Richter (2001), 
derived next to leading oder terms for the formfactor. Extensions to
all orders have been given by M\"uller \et. (2004). Gap conditions given 
by Tanner (2001) and Gnutzmann and Altland (2004) give lower and upper bounds 
for the border of universality.  For a recent review on quantum graphs see
Kuchment (2002).

Quantum dynamics on graphs became an issue also in the context 
of quantum information.  Aharonov \et (1993) 
pointed out that a random quantum walk on one dimensional chains can be faster
than the corresponding classical random walk. Since then, a whole field
has emerged dealing with quantum effects on graphs with properties superiour
to the corresponding classical operations. For an introductory overview 
and further references, see Kempe (2003).\\

We will in the following give a general definition of quantum graphs
and  discuss a specific set-up considered by Kottos and 
Smilansky (1997).  We will then review recent developments on the 
spectral statistics of quantum graph ensembles. Next, we discuss 
a special class of quantum graph 
ensembles, so-called regular quantum graphs. These types of graphs can 
show strong deviations from RMT depending on topological properties 
imposed on the graph in form of edge-colouring matrices. We will show that 
such graphs can be interpreted as realisations of quantum random walks 
on graphs.

\section{Quantum graphs - a brief review}
\subsection{Quantum graphs on line graphs} \label{sec:line}
In its most general form, a quantum graph is defined in terms of a (finite)
graph $G$ together with a unitary propagator $\bf U$; it describes the
dynamics of ''wavefunctions'' $\phi$ on the graph according to
\[ \phi_{n+1} = {\bf U} \phi_n \, , \]
such that waves can propagate only between connected vertices.
Motivated by physical application we will adopt a construction
of quantum graphs in terms of so-called line-graphs as explained below.

A (finite) \emph{directed graph}
or \emph{digraph} consists of a finite set of \emph{vertices} and a set of
ordered pairs of vertices called \emph{arcs}. We denote by $V^{G}$ and
$E^{G}$ the set of vertices and
arcs of the digraph $G$, respectively. Given an ordering of the
vertices, the \emph{adjacency matrix} of a digraph $G$ on $n$ vertices,
denoted by ${\bf A}^{G}$, is the $\left(  0,1\right)  $-matrix where
the $ij$-th element is defined by
\begin{equation}\label{adj-def}
A_{ij}^{G}:=\left\{
\begin{tabular}
[c]{ll}%
$1$ & if $(ij)\in E^{G},$\\
$0$ & otherwise.
\end{tabular} 
\right.
\end{equation}
An \emph{undirected graph} (for short, \emph{graph}) is a digraph whose
adjacency matrix is symmetric. The undirected connections between vertices
are called edges in this case. The \emph{line digraph} of a digraph $G$,
denoted by $LG$, is the graph which is obtained when taking the arcs 
as the new vertices; it is thus defined as  $V^{LG}=E^{G}$ and, given 
$(hi), (jk)\in E^{G}$, the ordered pair $((hi)(jk))\in E^{LG}$ if 
and only if $i=j$ (Bang-Jensen and Gutin 2001). 

\begin{figure}
  \begin{center} \includegraphics[height=4.0cm]{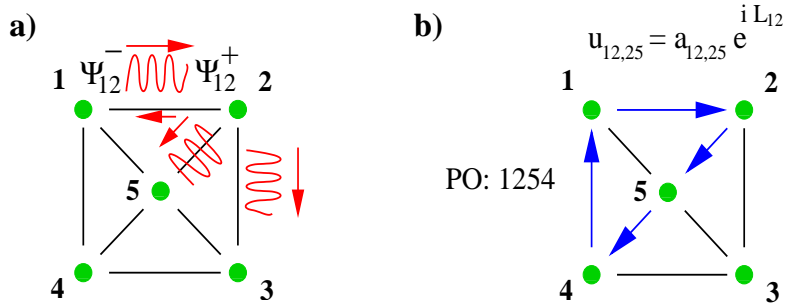} \end{center}
  \caption{a) Quantum graph; b) periodic path on a graph with quantum weights
defined with respect to the line - graph.}
  \label{Fig:graph}
\end{figure}

A \emph{quantum graph} associated with a digraph
$G$ on $n$ vertices can then be defined in terms of a set of unitary vertex
scattering matrices $\sigma^{(j)}$ on vertices $j=1,\ldots n$ and a set of
arc lengths $L_{ij}$ defined for every arc $(ij)\in E^{G}$. Waves
propagate freely along the directed arcs, transitions between incoming and
outgoing waves at a given vertex $j$ are described by the scattering matrix
$\sigma^{(j)}$, see Fig.\ \ref{Fig:graph}a. 
The two sets specify a unitary propagator of dimension
$n_{E}=|E^{G}|$ defining transitions between arcs
$(ij),(i^{\prime}j^{\prime})\in E^{G}$
which has the form (Kottos and Smilansky 1997)
\begin{equation} \label{SD}
\begin{tabular}
[c]{ccc}
${\bf S}^{G}={\bf D}\, {\bf V}$ & with & $D_{(ij)(i^{\prime}j^{\prime})}=
\delta _{i,i^{\prime}}\, \delta _{j,j^{\prime}}\, e^{\mathrm{i}kL_{ij}},$
\end{tabular}
\end{equation}
where $k$ is a wave number and
\begin{equation} \label{V}
\begin{tabular}
[c]{ccc}
$V_{(ij)(i^{\prime}j^{\prime})}=A_{(ij)(i^{\prime}j^{\prime})}^{LG}
\sigma_{ij^{\prime}}^{(j)}$ & with & $A_{(ij)(i^{\prime}j^{\prime})}
^{LG}=\delta_{j,i^{\prime}}$ $.$
\end{tabular}
\end{equation}
The local scattering matrices $\sigma^{(i)}$ describe the underlying
physical process which may be derived from boundary
conditions on the vertices. We will construct an
example below but may often regard the $\sigma^{(i)}$'s as arbitrary
unitaries. Let $d_{i}^{-}$ and $d_{i}^{+}$ be the number of incoming and
outgoing arcs of a vertex $i$, respectively. A sufficient and necessary
condition for a digraph $G$ to be quantisable in the way above is then, that
for every vertex $i\in V^{G}$, $d_{i}^{+}=d_{i}^{-} =
d_{i}=\dim\sigma^{(i)}$ (Pako{\'{n}}ski \et 2003). 
This means in particular that if $G$ is an undirected graph then it is 
quantisable.

Kottos and Smilansky (1997) considered solving the 1d Schr\"odinger equation
on an undirected graph assuming free propagation
on the arcs and imposing continuity and flux conservation at the vertices.
The solution on each arc propagating from vertex $i \to j$ takes on the form
\[ \phi(x_{ij}) = \phi_{ij}^{+} e^{i k x_{ij}} =
\phi_{ij}^- e^{-i k (L_{ij} - x_{ij})},\]
with $\phi^{\pm}_{ij}$ being the outgoing $(+)$ or incoming $(-)$ wave at
vertex $i$ or $j$. Continuity and flux conservation can then be written in
terms of the amplitudes $\phi_{ij}^{+}(0)$ and $\phi^-(L_{ij})$ at the 
vertices, that is

\begin{eqnarray*}
&\mbox{Continuity:}&
\phi^+_{ij} = \phi^-_{ji} = c_i \quad \mbox{for all $j$ with} (i,j) \in E^G\\
&\mbox{Flux cons.:}&
\sum_{j:(ij)\in E^G} \phi^+_{ij} = \sum_{j:(ji)\in E^G} \phi^-_{ji}
\end{eqnarray*}
These conditions give rise to local scattering matrices $\sigma^{l}$
mapping amplitudes $\phi_{il}^-$ onto $\phi_{lj}^+$ at vertex $l$
having the form
\begin{equation} \label{scat-matKS}
 \sigma^{(l)}_{ij} = -\delta_{ij} + \frac{2}{d_l}\,.
\end{equation}
The eigenvalue condition is then given as
\begin{equation} \label{spec-k}
\det({\bf I} - {\bf S}^G(k)) = 0
\end{equation}
with ${\bf S}^G(k) = {\bf D}(k) {\bf V}$ as defined in (\ref{SD}), (\ref{V}).
Scattering matrices for more general boundary conditions can be found
in Kottos and Smilansky (1999).

The "classical"\ dynamics corresponding to a quantum graph defined by a unitary
propagator ${\bf S}^{G}$ is given by a stochastic process with transition 
matrix $\bf T$ defined by
\begin{equation} \label{uni}
{\bf S}^G \to {\bf T}({\bf S}^G) \qquad \mbox{with} \quad
T({\bf S}^G)_{ij}=|S_{ij}^{G}|^{2}=|V_{ij}|^{2} \, .
\end{equation}
The matrix $\bf T$ is clearly stochastic, as $\sum_{j=1}^{n_E} T_{ij} = 1$ due
to the unitary of ${\bf S}^G$; the set of transition matrices related to a 
unitary
matrix as defined in (\ref{uni}) is a subset of the set of all stochastic
transition matrices, referred to as the set of {\em unitary-stochastic} 
matrices.  The topology of the set in the space of all stochastic matrices 
is in fact quite complicated, see Pako{\'n}ski \et (2001).  In what follows, 
we will only use that $\bf T$ has a largest eigenvalue 1 with corresponding 
eigenvector 
$\frac{1}{n_E}(1,\ldots,1)$ which follows from the Frobenius-Perron theorem 
and $\bf T$ being unitary-stochastic.

Note that both the quantum mechanics as well as the associated stochastic
dynamics relates to transitions between arcs and is thus defined on the line
digraph of $G$.

\subsection{Unitary stochastic ensembles and spectral statistics}

Inspired by the BGS conjecture for quantum
systems, we expect a link between the dynamical properties of the
stochastic process $\bf T$ and the statistical properties of the spectrum
of associated unitary matrices ${\bf S}^G$. It is thus natural to consider the
ensemble of unitary matrices
\begin{equation}
USE_{{\bf T}_0} = \left\{{\bf S}^G \mbox{is quantum graph on } G\,|
\;{\bf T}({\bf S}^G) = {\bf T}_0 \right \}
\end{equation}
for a given graph $G$ and a fixed unitary stochastic matrix ${\bf T}_0$ associated
with a stochastic process on the line-graph of $G$. Clearly, if 
${\bf S}^G(k) \in USE_{{\bf T}_0}$ for $k=0$ then it is for all $k\in \R$.
In fact, if all the arc length $L_{ij}$ are incommensurate, ${\bf D}(k)$ sweeps out
the space of unitary diagonal matrices of dimension $n_E$. For practical purposes
we will thus often replace averages over a given $USE_{\bf T}$ by averaging
over the space of diagonal unitary matrices
${\bf D}$ with $ D_{ij} = \delta_{ij} e^{{\rm i} \varphi_j}$
using the Euclidean measure on the $n_E$ - torus.

Rather than looking at the spectrum obtained from the secular determinant
(\ref{spec-k}), we will here consider the spectrum ${\bf S}^G$ for fixed
wavenumber $k$ and than average over $k$. One can write the spectrum 
in terms of a periodic orbit
trace formula reminiscent to the celebrate Gutzwiller trace formula being
a semiclassical approximation of the trace of the Green function
(Gutzwiller 1990).
We write the density of states in terms of the traces of ${\bf S}^G$, that is,
\be{density}
d(\theta) = \sum_{i=1}^{n_E} \delta(\theta - \theta_i)
= \frac{n_E}{2\pi} + \frac{1}{\pi}{Re}\sum_{n=1}^{\infty}
\Tr ({\bf S}^G)^n e^{-i n\theta} \, ,
\ee
where $\{\theta_i\}_{i=1,{n_E}}$ refers to the eigenphases of ${\bf S}^G$.
The traces $\Tr ({\bf S}^G)^n$ can be given as sum over
all periodic paths of length $n$ on the graph, i.e.\
\[
\Tr ({\bf S}^G)^n = \sum_{p}^{(n)} A_{p} e^{i k L_{p}}.
\]
Describing a given periodic path in terms of its vertex code
$(v_1, v_2 \ldots v_n)$, $v_i \in \{1,2,\ldots n\}$ with
$(v_i,v_{i+1})\in E^G$ being an allowed transitions between vertices, one
obtains for the amplitudes $A_p$ and lengths $L_p$ 
\be{weights}
A_p = \prod_{i=1}^n V_{(v_iv_{i+1}), (v_{i+1}v_{i=2})}, \qquad
L_p = \sum_{i=1}^n L_{v_iv_{i+1}}.
\ee
where the amplitudes give again transitions between arcs (not vertices).
An example of such a periodic path is given in Fig.\ \ref{Fig:graph}b.

Trace formulas like (\ref{density}) are a starting point for analysing the
statistical properties of quantum spectra. The statistical quantities such
as the two-point correlation function can be written in terms of the
density of states $d(\theta,N)$, that is,
\be{two-point1}
R_2(x) = \frac{1}{\overline{d}^2}
<d(\theta) d(\theta + x/\overline{d})>_{USE_{\bf T}, \theta} \, ,
\ee
where $\overline{d} = N/2\pi$ is the mean level density. The average
is taken here over the angle $\theta$ as well as over the USE (which is
equivalent to energy averaging). The Fourier coefficients of (\ref{two-point1})
can be written in terms of the traces of $\bf S$; one obtains
\be{K1}
K(\tau) = <\frac{1}{N} |\Tr {\bf S}^{N\tau}|^2 >_{USE_{\bf T}}
\ee
with $\tau = n/N$ and the average is taken over a USE. The so-called
form factor $K(\tau)$ can thus be written as a double sum over periodic
paths on the graph
\begin{eqnarray} \label{k_po1}
K(\tau) &=& <\frac{1}{N} \sum^{(n)}_{p,p'} A_p A_{p'}
e^{{\rm i} (L_p - L_{p'})} >_{USE_{\bf T}}\\
\label{k_po2}
 &\approx& g \frac{n}{N} \Tr {\bf T}^n  + <\sum^{(n)}_{p \ne p'} A_p A_{p'}
e^{{\rm i} (L_p - L_{p'})} >_{USE_{\bf T}}\,.
\end{eqnarray}
The first term in (\ref{k_po2}), also called the diagonal term (Berry 1985),
originates from periodic orbit pairs ($p, p'$) related through cyclic
permutations of the vertex symbol code.  There are typically $n$ orbits
of that kind and all these orbits have the same amplitude $A$ and phase $L$.
The corresponding periodic orbit pair contributions is (in general) $g\cdot n$ -
times degenerate where $n$ is the length of the orbit and $g$ is a symmetry 
factor ($g=2$ for time reversal symmetry). 

Expanding the random matrix result for the formfactor, one obtains for
$0 \le \tau = n/N \le 1 $
\begin{eqnarray*}
K_{CUE}(\tau) &=& \tau \\[.2cm]
K_{COE}(\tau) &=& 2 \tau - \tau \log(1-2 \tau)\\
              &=& 2 \tau - 2 \tau^2 + 2 \tau^3 + \ldots\, .
\end{eqnarray*}

\begin{figure}
  \begin{center} \includegraphics[height=3.0cm]{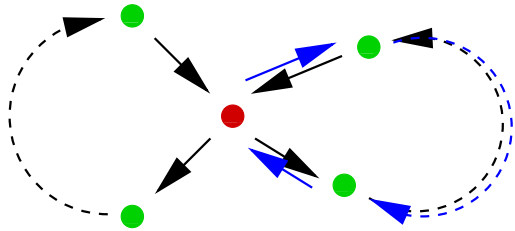} \end{center}
  \caption{Periodic orbit with one intersection on undirected graphs:
  two different paths exist having the same length following either
  the black or blue orbit.}
  \label{Fig:fig8}
\end{figure}

The linear terms are reproduced by the diagonal contribution which gives
the important link between the stochastic dynamics and the spectral
statistics; it follows from $\lim_{n\to\infty} \Tr {\bf T}^n = 1$
given there is a gap between the leading and next-leading eigenvalue
of $\bf T$. Contributions to the double sum in (\ref{k_po2}) which survive the
ensemble average are due to periodic orbit pairs where orbits visit
the same arcs along its path but in different order. The simplest examples
are ''figure eight'' orbits of the type shown in Fig.\ \ref{Fig:fig8} 
which arise in undirected graphs. Berkolaiko \et (2002) could indeed show that
orbits of that type give the correct ${\cal O}({\tau}^2)$ - contributions to
the GOE - form factor. In fact, a general scheme emerges relating
${\cal O}({\tau}^n)$ contributions to periodic orbits with $n$ intersections.
This work was inspired by a similar analysis for general quantum systems
by Sieber and Richter (2001). The periodic orbit contributions
giving the ${\cal O}(\tau^3)$ have been worked out by Berkolaiko \et (2003) 
and a general scheme for obtaining higher order terms iteratively has been 
given by M\"uller \et (2004).

\subsection{Border of universality}
\label{sec:border}
In the light of these recent developments, it becomes important to establish
the boundaries at which the spectral statistics of quantum graph
ensembles (or more general quantum systems) starts to deviate from 
random matrix behaviour. One can distinguish two different
scenarios. Firstly, the properties of the underlying dynamics, that is,
the stochastic process $\bf T$ may be linked to the spectral statistics
and may thus provide conditions for the onset of deviations from RMT
behaviour; such an approach is in the spirit of the original BGS -
conjecture making a connection between classical chaos and
random matrix statistics. Secondly, one may consider special phase or
length correlations in the quantum graph which could lead to interesting
non-universal statistics; in this approach, quantum graph ensemble
averages are carried out only over subsets of the full $USE$.

We will 
discuss known bounds on the border of universality related to the properties
of the stochastic process. An interesting family of quantum graphs which
belongs to the second category are so-called regular quantum graphs, which 
will be treated in more detail in section \ref{sec:reg-graph}.

\paragraph{Spectral gap conditions}
When studying the border of universality, we always need to consider the
limit of large graphs, that is, $n_E \to \infty$. This limit is in general
not well defined, but may often be obvious from the examples considered.
We will thus define the semiclassical limit loosely via a family of
unitary-stochastic transition matrices $\{{\bf T}_{n_E} \}$  and associate
$USE$'s and take $n_E \to \infty$. The leading term in (\ref{k_po2})
then gives a condition for a family to show deviations from RMT statistics 
in terms of the spectrum of $\bf T$; the diagonal term 
must obey 
\[
 1 - \Tr {\bf T}^{n_E \tau} \approx e^{-\Delta n_E \tau} \to 0 ;
\quad \mbox{for} \quad n_E \to \infty
\]
in order to match the leading coefficient in the expansion of the 
RMT form - factor; here $\Delta$ is the spectral gap, that is, 
$\Delta = -\log(1-|\Lambda_1|)$ with $\Lambda_1$ being the next to 
leading eigenvalue of ${\bf T}_{n_{E}}$) and $\tau = n/n_E$ is fixed. 
The condition above implies that we expect to
see deviations from RMT behaviour to leading order if 
\[ \Delta \sim n_E ^{-\alpha} \quad \mbox{with} \quad \alpha > 1, \]
that is, whenever the gap closes faster than $1/n_E$ for large system 
sizes (Tanner 2001). Based on super-symmetric techniques, Gnutzmann and 
Altland (2004) could give a lower bound by showing that the 
spectral gap condition guarantees RMT behaviour for  $\alpha \le 1/2$. The 
border of universality must therefore lie in the range $1/2 < \alpha \le 1$.
Other bounds have been given by Berkolaiko \et (2002,2003) which have been 
derived from higher
order terms in the expansion of the form factor. Below we we will give two
examples families with $\alpha = 1$ displaying critical behaviour by neither
converging to RMT nor to Poisson statistics in the semiclassical limit.

\begin{figure}
  \begin{center} \includegraphics[height=3.5cm]{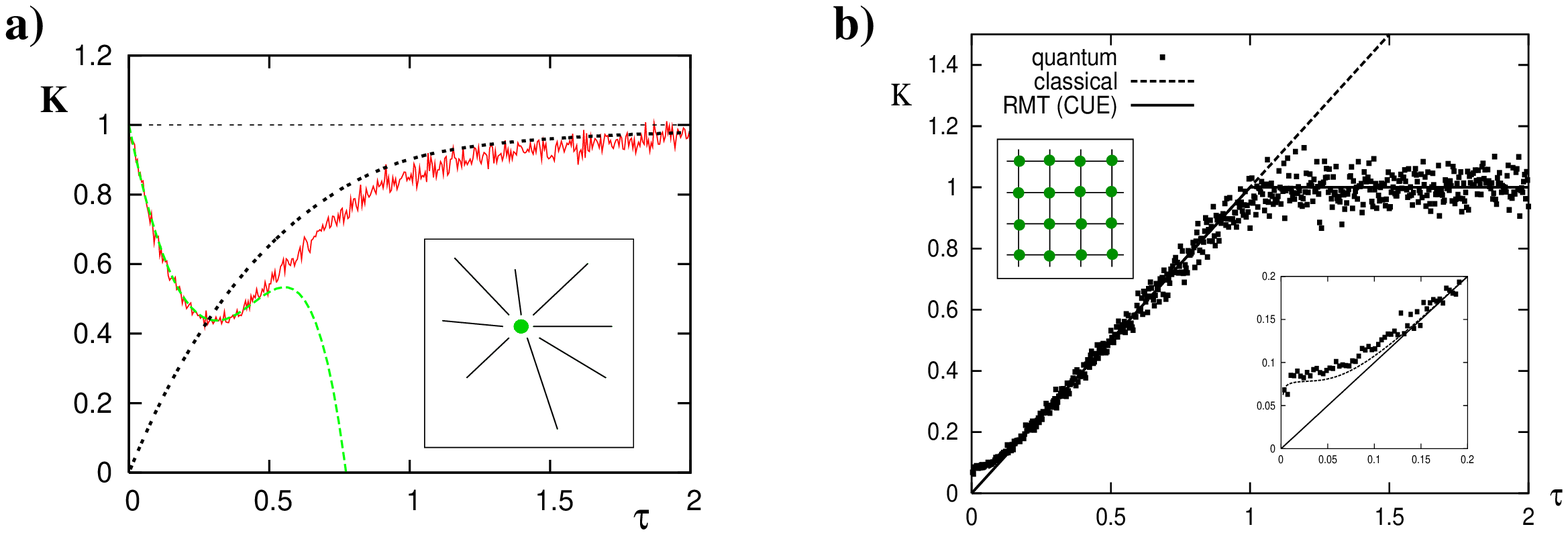} \end{center}
  \caption{Form factor for a) star graphs: numerics (red) versus the 
power series expansion (\ref{K-star}) (green); 
b) diffusive network (in 2 d): deviations occur for small $\tau$, see inset.
  }
  \label{Fig:graphdev}
\end{figure}

\paragraph{Quantum star-graphs:}
Quantum star-graphs arise naturally when one quantises a graph with
a single central vertex attached to $n_E$ undirected edges,
see inset of Fig.\ \ref{Fig:graphdev}a. The underlying graph is complete,
that is, we can reach every edge from every other edge through the central
vertex (ignoring the trivial dynamics on the outer vertices).
Typical boundary conditions imposed on the wave equation at the central
vertex following Kottos and Smilansky's approach in Sec.\ \ref{sec:line}
yield scattering matrices which greatly favour backscattering. The vertex
scattering matrix (\ref{scat-matKS}) which is essentially equivalent to the
matrix $\bf S$ here is
\begin{equation}  \label{u-star-qu}
S_{ij} = - \delta_{ij}  + \frac{2}{n_E}\; .
\end{equation}
The Markov processes associated with quantum star graphs correspond to  
systems of weakly coupled edges. Its dynamical properties are determined by 
the spectrum of the stochastic matrix associated with (\ref{u-star-qu}) which 
is highly degenerate and can be given explicitly (Kottos and Smilansky 1999), 
that is,
\[ \Lambda_0= 1, \quad \Lambda_1, \ldots, \Lambda_{n_E-1} =
(1 - \frac{4}{n_E}) \approx \frac{4}{n_E}\; .\]
Quantum star-graphs have therefore a critical classical spectrum with a
spectral gap vanishing proportional to $1/n_E$; one finds indeed
spectral statistics intermediate between Poisson and  COE  statistics.

The two-point correlation function has been work out explicitly
by Berkolaiko \et (2001) and has been shown to coincide with the
statistics of so-called Seba billiards, that is, rectangular billiards with
a single flux line. The first few terms in a power series expansion
of the form factor have been derived by Kottos and Smilansky (1999) and
Berkolaiko and Keating (1999) and yield
\begin{equation}  \label{K-star}
K(\tau) =  e^{-4\tau} + 8 \tau^3 \frac{32}{3}\tau^4 +
\frac{16}{3} \tau^5  + \ldots ,
\end{equation}
see Fig.\ \ref{Fig:graphdev}a.

\paragraph{Diffusive networks:}
The quantum mechanics of classically diffusive systems has been studied 
mainly in the context of Anderson localisation and 
localisation-delocalisation transitions, see e.g. Dittrich (1996)
and Janssen (1998) for recent review articles.

As a simple example, we consider here a quantum graph corresponding to 
a classical Markov process on a regular lattice in $d$ dimensions,
see Fig.\ \ref{Fig:graphdev}b for $d=2$.
Choosing a stochastic matrix ${\bf T}_d$ with constant transition 
probabilities $t_{ij} = 1/2 d$ between connected edges corresponds to   
$d$-dimensional diffusion in the continuum limit $L\to \infty$; here,
$L$ is the number of vertices along each direction, that is, the
total number of arcs is $n_E = 2 d L^d$.
Solving the diffusion equation with periodic boundary conditions allows
one to recover the low lying part of the spectrum of ${\bf T}_d$, that is, 
\be{spec-diff}
\log \Lambda_{\bf m} = - \frac{4 \pi^2 D}{L^2} \sum_{i=1}^d m_i^2,
\ee
with diffusion constant $D = \frac{1}{2 d}$ and $\bf{m}$ is a $d$-dimensional 
integer lattice vector. The influence of the dimension $d$ on the small 
$\tau$ behaviour of the form factor in diffusive systems has been described 
in detail by, for example, Dittrich (1996) and references therein. 
In terms of the spectral gap condition, one obtains
\[\Delta_N = \frac{4 \pi^2 D}{L^2} = \frac{4 \pi^2 (2 d)^{2/d-1}}{n_E^{2/d}} \]
that is, we expect universal statistics for $d\ge 3$ only. (Actually, 
the bound by Gnutzmann and Altland (2004) guarantees random matrix statistics 
only for $d \ge 4$, numerics suggests however, that $d=3$ follows 
RMT already (Tanner 2001)). One finds 
convergence to the Poisson limit due to Anderson localisation in one dimension 
(Schanz and Smilansky 2000). The two-dimensional case is critical with
$\Delta \sim 1/n_E$ which shows up in the form factor as a plateau 
for $\tau \to 0$, that is, $\lim_{\tau\to 0} K(\tau) = 1/4 \pi $, 
see Fig.\ \ref{Fig:graphdev}b (Tanner 2002).

\section{Regular quantum graphs}\label{sec:reg-graph}
 
In the following, we will consider quantum graphs for which the above 
bounds do not necessarily hold due to length correlations in the 
graphs, that is, averages are not taken over a full $USE$. We will 
actually look at a specific set of such non-generic quantum graphs, 
namely quantum graphs for which the global propagator $\bf S$ consists of 
identical local scattering processes at every vertex. Such graphs have 
been called {\em regular quantum graphs} by Severini and Tanner (2004); 
the name derives from the notation {\em regular graphs} for graphs which 
have the same number of 
incoming and outgoing arcs at every vertex and thus share the property
that vertices are locally indistinguishable. We will show that
regular quantum graphs corresponding to the same underlying graph $G$ can 
behave very differently depending on how the local scattering
processes are connected to each other. We will show that a crucial 
element in this is played by the possible ways regular graphs can be 
edge-coloured. Regular graphs are in fact another way at looking at 
{\em quantum random walks} as will be pointed out at the end of this 
section.

\subsection{Regular quantum graphs and edge-colouring matrices}
\begin{figure}
  \begin{center} \includegraphics[height=3.5cm]{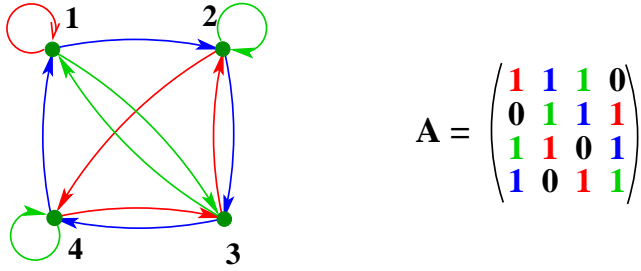} \end{center}
  \caption{A 3-regular graph of size 4 together with a possible
   edge-colouring; the edge-colouring matrices correspond to entries having
   the same colour in the adjacency matrix $A$ of the graph.}
  \label{Fig:reggraph}
\end{figure}

We will construct a quantum graph on a $d$-regular digraph $G$ with 
$n$ vertices for which the wave dynamics at a given vertex of the 
graph is "locally indistinguishable" from that of any other vertex 
of the graph. This is
done by choosing a unitary $d$-dim.\ matrix $\sigma$ and a set of
arc-lengths $L_i, i=1,\ldots d$ and ascribing the scattering process 
$\sigma$ to every vertex in the graph with incoming as well as outgoing
arcs chosen from the set of $L_i$'s at every vertex.
This is done here by first fixing a so-called {\em edge-colouring} of 
the graph (see eg Bollob\'as 1979), that is, we assign one of $d$ 
different ''colours'' to every directed arc of the
graph in such a way that no vertex has two incoming or two outgoing arcs
of the same colour. Note that there are many different ways to 
edge-colour a given regular graph for $d \ge 2$.  
Edge-colouring can be described in terms of a set of 
$n$ - dimensional permutation matrices $\rho_i, i=1,\ldots d$ having the 
property that
\begin{equation}
\sum_{i=1}^{d}\rho_{i}={\bf A}^{G} \, 
\label{cond-rho}
\end{equation}
with ${\bf A}^G$ being the adjacency matrix of $G$;
the arc $(ij)$ is then assigned the colour $k$ if the $(ij)$ matrix
element of $\rho_k$ is non-zero. We refer to the set of $\rho_i$'s as
the {\em edge-colouring matrices} of $G$\footnote{In Severini and 
Tanner (2004), these matrices have been called {\em connectivity matrices}.}. 
In Fig.\ \ref{Fig:reggraph}, an explicit example is shown with 
\[ %\begin{eqnarray*}
{\bf A}=
\left(\begin{array}{cccc}1&1&1&0\\0&1&1&1\\1&1&0&1\\1&0&1&1\end{array}\right),
\quad
\rho_1=
\left(\begin{array}{cccc}1&0&0&0\\0&0&0&1\\0&1&0&0\\0&0&1&0\end{array}\right),\;
\rho_2 = 
\left(\begin{array}{cccc}0&1&0&0\\0&0&1&0\\0&0&0&1\\1&0&0&0\end{array}\right),\;
\rho_3 = 
\left(\begin{array}{cccc}0&0&1&0\\0&1&0&0\\1&0&0&0\\0&0&0&1\end{array}\right).
\] %\end{eqnarray*}

The adjacency matrix ${\bf A}^{LG}$ of the line-graph of $G$ can then be written 
in the form (Severini 2003, Severini and Tanner 2004)
\begin{equation}
{\bf A}^{LG}=\left(  \bigoplus_{i=1}^{d}\rho_{i}\right)  
\cdot\left( {\bf J}_{d}\otimes
{\bf I}_{n}\right)  \text{ }\, , \label{adj-LG}
\end{equation}
where ${\bf J}_{d}$ is the $(d\times d)$ matrix with all elements being equal 
to 1 and ${\bf I}_{n}$ is the identity matrix. We proceed by defining a 
quantum graph on the line-graph of $G$ in the form of a unitary propagator
${\bf S}^G$ as follows:
\begin{equation}
{\bf S}^{G}=\left(  \bigoplus_{i=1}^{d}\rho_{i}\right)  
\cdot\left( {\bf  C}\otimes {\bf I}_{n}\right)  =\left[
\begin{array}
[c]{cccc}
C_{11}\,\rho_{1} & C_{12}\,\rho_{1} & \cdots & C_{1d}\,\rho_{1}\\
C_{21}\,\rho_{2} & C_{22}\,\rho_{2} & \cdots & C_{2d}\,\rho_{2}\\
\vdots & \vdots & \ddots & \vdots\\
C_{d1}\,\rho_{d} & C_{d2}\,\rho_{d} & \cdots & C_{dd}\,\rho_{d}%
\end{array}
\right]  , \label{qu-LG}%
\end{equation}
with ${\bf C}(k) = {\bf D}(k)\,\sigma$ and 
$D(k)_{il} = \delta_{jl} e^{i k L_j}$ describing the local 
scattering process. The $d$ - dimensional matrix $\bf C$ is 
also called the \emph{coin} in the context of quantum random walks on graphs 
(Kempe 2003). 

Note that different ways of edge-colouring the graph, that is, different 
decomposition of $A^{G}$ in the form (\ref{cond-rho}) lead to 
different quantum graphs which may have quite different properties 
as will be shown in the next section; this is in 
contrast to the representations of the line-graph adjacency matrix 
(\ref{adj-LG}) which are all equivalent up to relabeling the arcs in 
the graph. For modifications of this construction for undirected graphs
with time reversal symmetry, see Severine and Tanner (2004).

\subsection{From 'integrable' to 'chaotic' regular quantum graphs - 
some examples}
\begin{figure}
  \begin{center} \includegraphics[height=3.5cm]{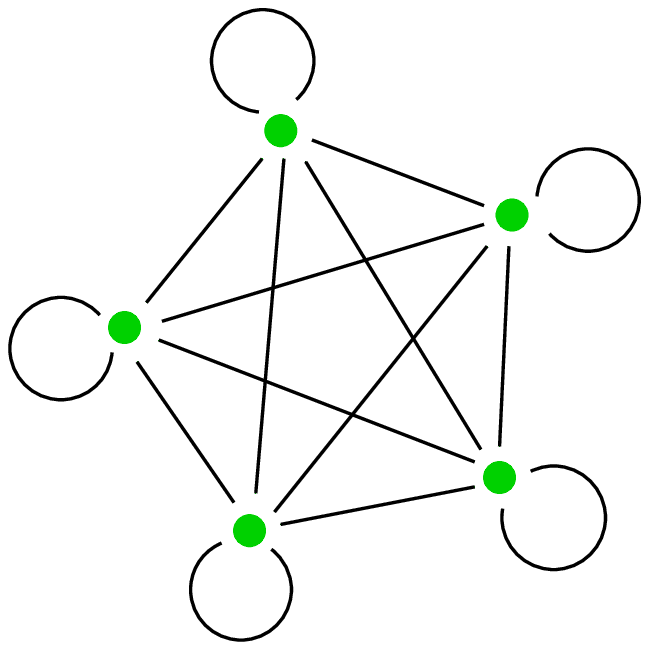} \end{center}
  \caption{The complete graph $K^n$ with $n=5$.}
  \label{Fig:complete}
\end{figure}

In this section, we will show that different ways to edge-colour a graph
can indeed lead to very different types of quantum graphs with spectral
statistics ranging from Poisson to CUE. In other words, in regular quantum 
graphs it is the choice of the edge-colouring matrices $\rho_i$, a purely 
topological quantity, which determines the properties of the quantum graph
independent of the single vertex scattering processes given through
the coin $\bf C$. We will demonstrate this here for a specific example, namely
so called \emph{complete graphs} $G = K^n$ with adjacency matrix ${\bf A}^G 
= {\bf J}_{n}$, that is every vertex is connected to every other vertex, 
see Fig.\ \ref{Fig:complete}. (Note that $n_E = n^2$, here). A more
general treatment can be found in Severini and Tanner (2004).
In the examples discussed below, we make use of the fact that for 
finite groups $\Gamma$ of order $n$, we may write 
\begin{equation}
\sum_{i=1}^{n}\rho_{i}={\bf J}_{n}\,. \label{cond-Jd}
\end{equation}
where the $\rho_i$'s form a regular representation of $\Gamma$.
In what follows we will study various decompositions of $J_{n}$ and see
how they effect statistical properties of the spectra of quantum graphs.

\subsection{The cyclic group $\mathbb{Z}_{n}$}

We will first consider an abelian group, namely the cyclic group 
$\mathbb{Z}_{n}$. The $\rho_{i}$'s forming a regular representation
commute with each other and are of the form
\[
\begin{tabular}
[c]{ccccc}
$(\rho_{j})_{kl}=\delta_{k,(l+j)\operatorname{mod}n}$ & with eigenvalues &
$\chi^j_{m}=e^{2\pi\mathrm{i} \frac{jm}{n}},$ & where & $j,m=1,\ldots, n\,.$%
\end{tabular}
\
\]
The abelian nature of the group allows one to block-diagonalise the matrix
${\bf S}^G$ into $n$ blocks of dimension $n$ each, independent of the coin 
$\bf C$. 
The spectrum of the quantum graph is then given by the spectra of the 
sub-matrices 
\[
{\bf S}_{m}^{G}=
\left( \bigoplus_{j=1}^n e^{2\pi\mathrm{i} \frac{j m}{n}} \right) \cdot {\bf C} 
\quad \mbox{with} \quad  m = 1,\ldots, n \, .
\]
The eigenvalues of ${\bf S}^G$ are here characterised in terms of two 
quantum numbers, an 'angular momentum' quantum number $m$ and a second 
quantum number counting the eigenvalues in each $m$ manifold. 
If the spectra for different $m$ are uncorrelated, one expects Poisson 
statistics of the total spectrum in the limit $n\rightarrow\infty$.
 
\begin{figure}
  \begin{center} \includegraphics[height=19cm]{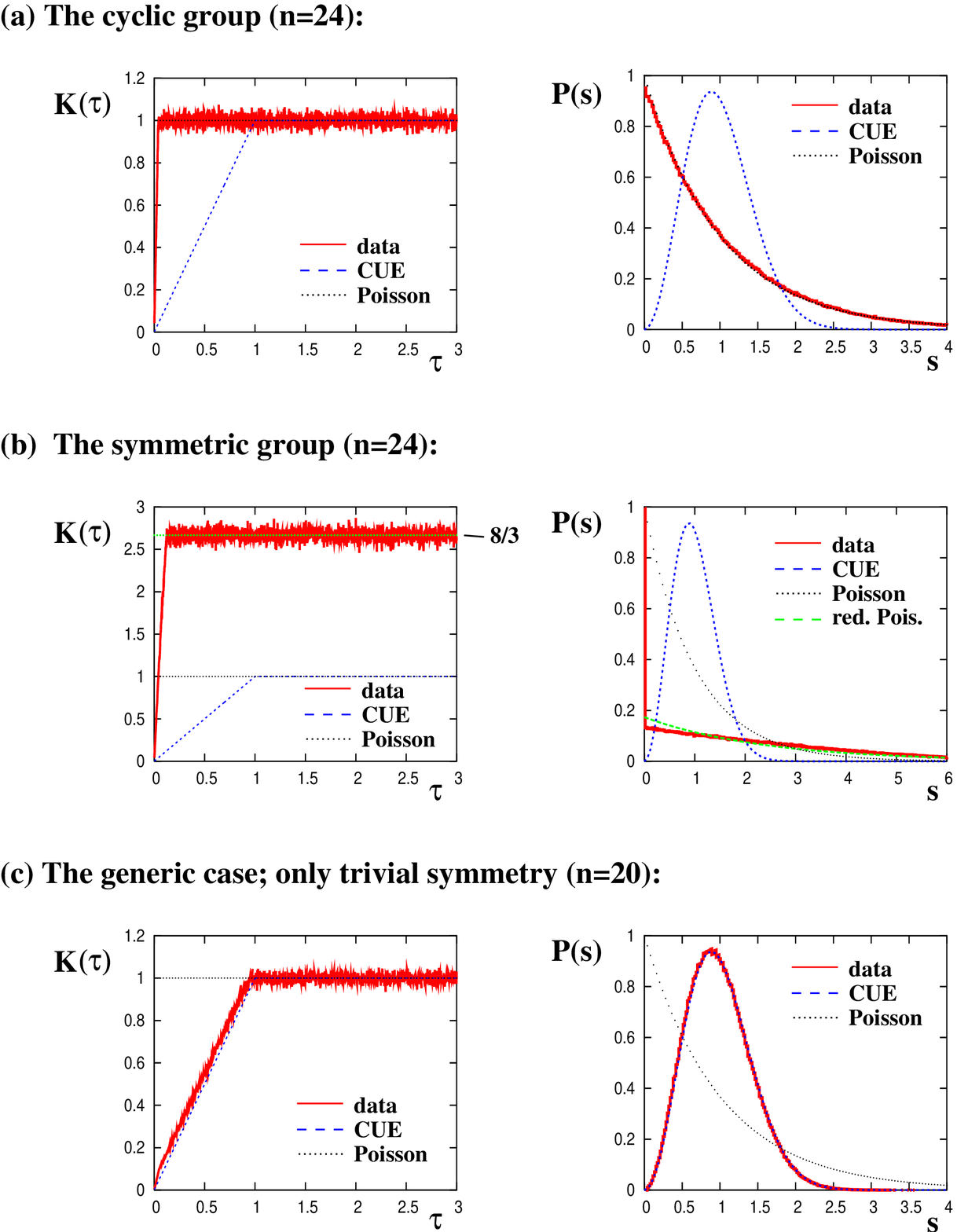} \end{center}
  \caption{Formfactor $K(\tau)$ and nearest neighbour spacing
  distribution $P(s)$ for: (a) the $\rho_i$'s form the regular
  representation of the cyclic group $\mathbb{Z}_{24}$; (b)
  $\rho_i$'s represent the symmetric group $S_4$; (c) a 'random' set
  of $\rho_i$'s without symmetries.
  The dashed curve in (b) labeled ''red.\ Poisson''
  corresponds to a distribution of degenerate levels being
  Poisson distributed otherwise.}
  \label{fig:stat}
\end{figure}

Figure \ref{fig:stat}a) shows spectral properties of ${\bf S}^{G}$ with $n=24$,
that is, $\dim S^{G}=576$. We plot here the nearest neighbour spacing (NNS)
distribution $P(s)$ and the form factor $K(\tau)$. 
The coin is of the from $C(k) = D(k)\, \sigma$ where the local scattering 
matrix $\sigma$ is choosen to be the Fourier matrix and the arc lengths
entering the diagonal matrix $D$ are chosen independently and identically
distributed in $[0,1]$. The average is, for a fixed choice of the coin, 
taken by averaging over the wavelength $k$. The numerical 
results are shown in Fig.~\ref{fig:stat}a) and
suggest indeed Poisson-statistics apart form deviations in the form factor 
on scales $\tau\leq1/n$ due to the `random nature' of the coin.

\subsection{The non-abelian case: the symmetric group $S_{4}$}
 
Next, we consider a specific example of a non-abelian group, namely the
symmetric group $S_{4}$ with $n=24$ elements. By writing  
the permutation matrices $\rho_i$ forming the regular representation 
of $S_{4}$ in terms of the irreducible representations (for
short \emph{irreps}) of $S_{4}$, we can again give the propagator 
$S^{K^{24}}$ in block-diagonal form where the blocks are now of size
$n\, d$, with $d$ being the dimension of the irrep under consideration.
The group $S_{4}$ has 2 one-dimensional, 1 two-dimensional and 
2 three-dimensional irreps, and each $d$ - dimensional irrep is 
contained $d$ times in each $\rho_i$ according to the 
general formula
\[
2\cdot1^{1}+1\cdot2^{2}+2\cdot3^{3}=24.
\]
We thus have 5 independent sub-spectra making up the spectrum of the 
quantum graph ${\bf S}^G$ of which two are of dimension $n = 24$, one is of 
dimension $2 \cdot 24 = 48$ and two are of dimension $3 \cdot 24 = 72$; the 
latter once are two and three times degenerate, respectively.  
The huge degeneracy in the spectra can clearly be seen in the
spectral statistics; it is manifest in the peak at $s=0$ in $P(s)$,
see Fig.\ \ref{fig:stat}b, and leads to
\[
\begin{tabular}
[c]{ccc}
$K(\tau)=(2\cdot3^{3}+1\cdot2^{3}+2\cdot1^{3})=8/3$ & for & $\tau>3/24\,.$%
\end{tabular}
\ \
\]
The spectra appear to be uncorrelated otherwise; note however, that the
spectrum for each sub-block alone are correlated following $CUE$
statistics, which gives rise to the deviations from purely Poisson
behaviour in $P(s)$ (cf. dashed curve) as well as in the behaviour of the form
factor for $\tau\leq 3/24$ which is dominated by the sub-spectra of the three
dimensional irreps.

\subsection{The generic case: no symmetries}

The overwhelming number of decompositions of the form (\ref{cond-Jd}) will of
course have no common symmetry, that is, it is not possible to 
block-diagonalise the $\rho_i$'s simultaneously. We therefore do not expect any 
special features in the spectrum. The question remains, however, if the 
'randomness' put into the system by choosing a random edge-colouring is 
enough to produce generic, random matrix type, statistics. After all, 
these quantum graphs still possess a large degree of degeneracy due to the
presence of identical coins at every vertex.
A numerical study may thus reveal interesting insights into the
range of validity of the RMT - regime. Fig.\ \ref{fig:stat}c
shows the level statistics for an unstructured choice of edge-colouring
matrices which is in good agreement with random matrix theory for the 
CUE - ensemble \footnote{Deviations in the formfactor for small $\tau$ can be
attributed to the fact that the spectrum of $C$ itself is always contained 
in the full spectrum, see Severini and Tanner (2004); 
this part has been removed in the NNS statistic.}. Note that the statistic 
has been obtained from the spectrum on an $(n^{2}\times n^{2})$ unitary 
matrix which has only $n^{3}$ non-zero elements of which only $n^{2}$ 
are independent; in addition, there are only $n$ different arcs lengths to 
choose from for $n^2$ different arcs. The origin of the universality in the 
spectral statistics in this type of quantum graphs is here clearly not due 
to the 'randomness' in the choice of the matrix elements but due to 
random  edge-colouring of the graph alone!

\subsection{Quantum random walks}
In recent years, the study of unitary propagation on graphs has also been 
looked at from the perspective of devising a quantum version of a 
random walk. This line of thought arose as part of
the effort to build quantum information systems being able to do operations 
which are impossible or much slower on classical devices. It could 
indeed be shown that quantum random walks have this property under certain
circumstances: quantum walks can be faster for some network geometries 
(Aharonov \et 1993) and can even lead to an exponential speed-up
such as for the graph-traversal algorithm by Childs \et (2003);
for an introductory overview and further references, see Kempe (2003). 
The generalisation of Grover's algorithm (Grover 1997) to spatial 
searches on graphs (Shenvi \et 2003, Ambianis \et 2004, Childs and 
Goldstone 2004) is another remarkable recent result.

We will not study the properties of quantum random walks here; 
instead, we would like to point out that the discrete quantum 
walk modules discussed in the literature are in fact equivalent to 
regular quantum graphs such as introduced in the previous
sections. 

In brief, a discrete quantum random walk comprises of an in general 
$d$ - regular graph $G$ with $n$ vertices (where lattices are favoured in the 
literature) and a 'spin' degree of freedom, where the spin can take up $d$ 
different states.  A quantum state at a vertex $v$ with spin $i$ is then 
described by $|v,i>$ and the quantum wave function is a superposition of 
all possible vertex states. The quantum random walk consists of a 
'walk' element and a 'quantum coin toss'. The walk is steered by the internal 
spin states, that is, there exist matrices $\rho_i$, $i=1,\ldots d$ which 
define in which direction the component $|v,i>$ is flowing, that is,
\[ \rho_i |v,i> = |v',i> \]
where the internal spin states remains unchanged. The $\rho_i$'s are 
obviously permutation matrices in the vertex space. The coin toss is
simulated by applying a unitary transformation $C$ to the spin states at
every vertex, that is,
\[{\bf C} |v,i> = |v, \sum_j C_{ij} j>;\]
the coin is unbiased, if $|C_{ij}|^2 = 1/d$ throughout. It is clear from the
construction, that one obtains a classical random walk on such a network, 
if one performs a measurement after every walk by projecting out
the spin degrees of freedom.

A full cycle of walk and coin toss is then described by a unitary matrix
\begin{equation}
{\bf S}_{QW}=\left(  \bigoplus_{i=1}^{d}\rho_{i}\right)  
\cdot\left(  {\bf C}\otimes {\bf I}_{n}\right)  
\end{equation}
which has exactly the form of the propagator for regular quantum graphs in 
Eq.\ (\ref{qu-LG}). We thus identify the $\rho_i$'s with the 
edge-colouring matrices and the coin is in fact a local scattering matrix. 
Writing the components of the wave function in $|v,i>$ notation is thus
just a different way of labeling the arcs in the graph and the dynamics
takes indeed place on the line-digraph of $G$.

To demonstrate the mechanism, let us give a specific example: the quantum  
walk on an infinite line. A popular setting is the following walk:
\begin{eqnarray*} 
\mbox{states:} 
&\quad& |i,\uparrow>,\; |i, \downarrow>, \quad i \in \mathbb{Z}\\[.2cm]
\mbox{walk:} &\quad& 
\begin{array}{ccc}
|i,\uparrow> &\to& |i+1, \uparrow> \\ |i,\downarrow>& \to& |i-1, \downarrow> 
\end{array},
\quad \mbox{that is, } \quad \rho_{\uparrow} = \delta_{i,i+1}, 
\quad \rho_{\downarrow} = \delta_{i,i-1}; \\[.2cm]
\mbox{coin:} &\quad& {\bf C} = \frac{1}{\sqrt{2}}
\left(\begin{array}{cc} 1&1\\ -1&1 \end{array}\right)\, . 
\end{eqnarray*}
A random walk starting in the state $|i,\uparrow>$ then evolves according to
\[
|i,\uparrow> \stackrel{C}{\to} \frac{1}{\sqrt{2}} 
\left(|i,\uparrow> - |i, \downarrow> \right) \stackrel{\mbox{walk}}{\to}
\left(|i+1,\uparrow> - |i-1, \downarrow> \right) \stackrel{C}{\to} \ldots \, .
\]
Continuing this process leads quickly to a probability profile which
grows linearly with the number of steps $t$ which is in contrast to the
classical spreading for 1d random walks being of the order $\sqrt{t}$
(Kempe 2003). It must be pointed out, however, that this is a pure symmetry
effect due the choice of edge-colouring matrices; in the setting 
chosen here, the two operations $\rho_{\uparrow}$ and $\rho_{\downarrow}$ 
commute. By introducing disorder into the quantum random walk on a line 
by for example randomly switching the shift operation at every vertex 
leads to a completely different behaviour resembling that of a quantum graph 
on a diffusive network as discussed in sec.\ \ref{sec:border}. \\

\noindent
\textbf{Acknowledgment:}

\noindent
The author would like to thank Simone Severine with whom some of the work
reviewed here has been carried out. Thanks goes also to the Royal Society and 
Hewlett-Packard, Bristol, for financial support.

\end{document}